\documentclass{article}

\usepackage{microtype}
\usepackage{graphicx}
\usepackage{subcaption}
\usepackage{booktabs} 
\usepackage{multirow}
\usepackage{tabularx}
\usepackage{xcolor}
\usepackage{listings}
\usepackage{inconsolata}
\usepackage{floatrow}

\usepackage{hyperref}

\lstdefinestyle{pseudo}
{
 language=python,
 linewidth=\columnwidth,
 stringstyle={\color{gray}},
 basicstyle=\scriptsize\ttfamily,
 keywordstyle=\color{black}\bfseries,
 morekeywords={abstract,class,function},
 morecomment=[l][\color{gray}]{//},
 numbers=left,
 xleftmargin=2em,
 mathescape=true,
}

\newcommand{\circled}[1]{\scalebox{1.1}{\raisebox{.4pt}{\textcircled{\raisebox{-.4pt} {\scalebox{.8}{#1}}}}}}

\usepackage[accepted]{icml2026}

\usepackage{amsmath}
\usepackage{amssymb}
\usepackage{mathtools}
\usepackage{amsthm}

\usepackage[capitalize,noabbrev]{cleveref}

\newcommand{\op}[1]{\texttt{\small #1}}
\theoremstyle{plain}

\theoremstyle{definition}

\theoremstyle{remark}

\newtheorem{example}{Example}

\icmltitlerunning{Evaluating Next Action Prediction Systems for Spreadsheets}

\begin{document}

\twocolumn[
  \icmltitle{A Benchmark and Framework for Evaluating Next Action Predictions in Spreadsheets} 



  \icmlsetsymbol{equal}{*}

  \begin{icmlauthorlist}
    \icmlauthor{Tejas Agrawal}{msft_i}
    \icmlauthor{Vu Le}{msft}
    \icmlauthor{Sumit Gulwani}{msft}
    \icmlauthor{Gust Verbruggen}{msft_b}
  \end{icmlauthorlist}

  \icmlaffiliation{msft_i}{Microsoft, Bangalore, India}
  \icmlaffiliation{msft}{Microsoft, Redmond, USA}
  \icmlaffiliation{msft_b}{Microsoft, Keerbergen, Belgium}

  \icmlcorrespondingauthor{Tejas Agrawal}{t-tagrawal@microsoft.com}
  \icmlcorrespondingauthor{Gust Verbruggen}{gverbruggen@microsoft.com}

  \icmlkeywords{Machine Learning, ICML}

  \vskip 0.3in
]



\printAffiliationsAndNotice{}  

\begin{abstract}
Predictive code completion greatly accelerates how quickly developers work.
In spreadsheets, despite being much more common, such auto-completion features are virtually non-existent.
To address this gap, we introduce a benchmark for systems that observe a sequence of user actions in a spreadsheet and predict future actions.
Two challenges are (1) the absence of edit histories in public spreadsheet corpora and (2) the complex space of spreadsheet actions (spatial, temporal, composite).
To address (1), we manually curate 52 sequences of 12K actions that recreate spreadsheets from public corpora, seeded by parametrized heuristics and LLM refinement.
To address (2), we propose an online evaluation that expects a prediction after each user action, accepts or rejects that prediction, updates the future actions upon acceptance, and repeats this until the target spreadsheet is obtained.
We use multiple baseline predictors (including zero-shot LLMs, fine-tuned SLMs, and classical models) and analyze different properties that our benchmark teaches us, including but not limited to: properties of saved actions and false positives, efficiency, effect of user profiles, effect of triggers, and effect of context.

\end{abstract}

\section{Introduction}
\label{sec:intro}

\begin{figure}[htb]
    \centering
    \includegraphics[width=.85\linewidth]{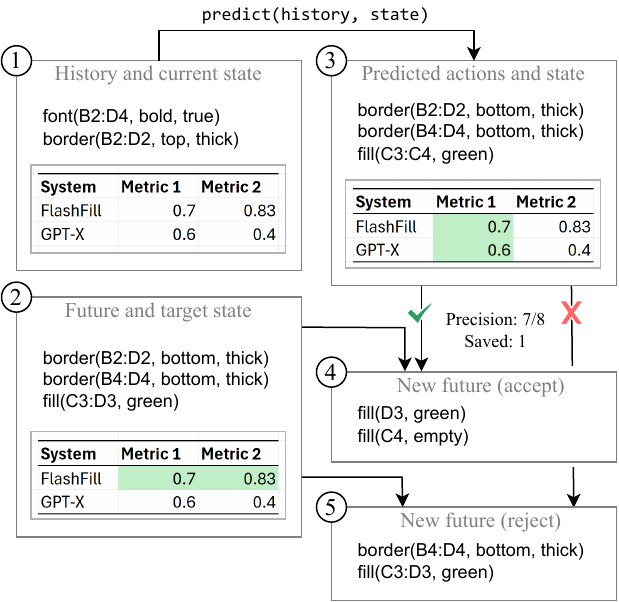}
    \caption{One step of the online evaluation on a synthetic example.
    \circled{1} The history (so far) produces the current state.
    \circled{2} The remaining future actions would produce the target state.
    \circled{3} Given the history and current state, the system predicts three actions:
    two borders that match the future, and a \textsf{\small fill} that targets the wrong cell.
    At the (cell,\,property) level this gives precision $7/8$;
    if accepted, the user would save one action overall (see below).
    The system then either \circled{4} \emph{accepts} or \circled{5} \emph{rejects} the prediction.
    On accept, the future is updated to remove operations the prediction already
    satisfied (the two borders) and a new \textsf{\small fill(C4, empty)} is prepended
    to undo the wrongly-coloured cell, leaving two actions instead of three.
    On reject, the future is unchanged and evaluation continues with the next user step.}
    
    \label{fig:introExample}
\end{figure}

Creating visually interesting spreadsheets requires typically hundreds of simple UI actions like selecting relevant ranges and clicking buttons to enter data or apply formatting.
We introduce the first benchmarking dataset and framework for predictive auto-completion in spreadsheets, where the system observes the user performing a sequence of these low-level actions and then suggests the next actions to speed up the spreadsheet authoring process.
This work thus bridges the gap between two emerging areas: code auto-completion and the rapid rise in developer productivity, and spreadsheet productivity and the rise of agents for spreadsheet authoring.

Code auto-completion has drastically evolved in recent years.
Early symbolic systems would observe users making changes and use small programs to suggest repetitive edits to the rest of the file, like Blue-Pencil \cite{miltner2019fly}.
With the rise of transformers architectures, full lines of code could now be suggested from potentially noisy existing code \cite{svyatkovskiy2020intellicode}.
As models became more powerful, so would their suggestions: modern coding assistants can suggest multiple functions in a pure auto-completion setting \cite{mastropaolo2023robustness}.

In spreadsheets, early assistants would auto-complete columns that could be derived from other columns, most notably FlashFill \cite{gulwani2011automating,cambronero2023flashfill++}.
Other assistants suggest formulas after the user explicitly signals the intent that they want a formula \cite{chen2021spreadsheetcoder}.
Because of the complex interaction surface of spreadsheets, two dimensions of cells with stylistic properties, spreadsheet suggestions are constrained to very specific scenarios.
More recently, the main focus for spreadsheet suggestions are agents that solve problems based on natural language intent \cite{ma2024spreadsheetbench,li2023sheetcopilot}.
For routine edits and repetitive patterns, the effort of invoking the assistive tool, formulating prompts, and waiting for responses exceeds the cost of direct manipulation, leading users to default to manual editing \cite{vaithilingam2022expectation,liang2025tabletalk}.

To the best of our knowledge, there is no work on generalized action suggestions for spreadsheets.
One potential reason is that evaluating predictive assistants in spreadsheets is challenging for various reasons.
First, as opposed to detailed version histories for code, there only exist very high-level corpora of spreadsheets as they evolve over time \cite{dou2016venron}.
Second, whereas each individual action (like coloring a cell or adding a border) is simple, different actions happen at (and affect) different parts of the spreadsheet in potentially many different orders, without a clear trigger of when to make a prediction.
A potential solution is teacher-forced offline evaluation, where the system is always given the correct next state and each prediction is checked independently against a fixed target.

We address the first challenge by manually curating sequences of user actions from static workbooks, using a combination of heuristics and large language models for heuristic seeding to motivate diversity.
Our dataset consists of 52 high-quality trajectories of workbooks being created from start to finish, each having between 35 and 821 steps with an average of 229 steps, for 11,907 steps in total.

We address the second challenge by proposing an online evaluation, instead of a static \emph{``given x, predict y''}.
At each iteration, the system to be evaluated is given the past $n$ actions and predicts a sequence of zero or more actions.
Based on properties of the prediction, that prediction is either accepted or rejected.
If accepted, the sequence of future actions is updated, for example, to remove successful actions and add actions that undo wrong moves.
If rejected, the next user action is added to the history and the next iteration starts with $n + 1$ actions.
This procedure is repeated until either there are no actions left to do (final spreadsheet obtained) or a threshold is reached.
Our online evaluation has three advantages over a teacher-forced offline evaluation at fixed time steps: (1) suggestions change the state that the model (and user) continues from, so errors can compound and later predictions depend on earlier ones; (2) the model cannot re-suggest the same easy actions and still appear accurate under stepwise scoring; and (3) the system is tested on revising or repairing its own prior suggestions.

\begin{example}
An example of a single iteration is shown in Figure~\ref{fig:introExample}.
In this hypothetical prediction, the two \textsf{\small border} actions are correct and the \textsf{\small fill} is partially correct, for a total (per-cell) precision of 7/8.
Because one correction operation is added to the new future to remove the wrong fill, the user saves one action.
\end{example}

Our framework then computes different metrics at different levels (per simulation, per prediction, per action) that allow targeted improvement of action prediction systems.
We use multiple solvers as baseline prediction systems (such as zero-shot LLMs, finetuned SLMs, and some classical machine learning models) to highlight what we can learn using these metrics.
For example, more powerful models save more actions than weaker models (33\% on GPT-5 with reasoning versus 18\% on GPT-5 mini) and fine-tuned 360M-parameter models match GPT-5 (both at 27\%).
This indicates that the task is learnable.
Low-precision acceptance heuristics yield negative savings ($-$19\%), confirming that abstention based on net user benefit is crucial.
Being able to predict at each action saves more actions (27\%) than after every four actions (17\%), indicating that good triggers or cheaper models are worth investigating.
Section~\ref{sec:experiments} describes many more insights.

In summary, we make the following contributions 
\begin{itemize}
    \item \textbf{Benchmark dataset.} We curate a dataset of 52 spreadsheet creation trajectories totaling 11,907 operations, each one validated by humans.
    \item \textbf{Online evaluation.} We propose an evaluation framework that evaluates systems end-to-end by modeling user acceptance behavior, dynamically adapting ground-truth trajectories after predictions, and computing metrics that capture real-world utility.
    \item \textbf{Design insights.} We demonstrate our dataset and framework using baselines to show how these metrics allow targeted improvement of predictive assistants for spreadsheets.
\end{itemize}
and release artifacts at {\small{\url{www.github.com/Tej-55/NAPE}}}.

\section{Benchmarks}
\label{sec:benchmark}

Each benchmark consists of a sequence of parametrized actions that a user takes to build a final workbook from scratch.
Table~\ref{tab:actions} lists the supported actions.

\begin{table}[]
    \centering
    \small
    \begin{tabularx}{\columnwidth}{lX}
    \toprule
    \textbf{Operation} & \textbf{Explanation} \\ \midrule
    \textsf{input(d, v)}        & Enter \textsf{v} (value or matrix) at location \textsf{d}.\\
    \textsf{merge(d)} & Merge range \textsf{d}.\\
    \textsf{format(d, f)}       & Set number format \textsf{f} at range \textsf{d}. \\
    \textsf{fill(d, c)}         & Set background fill color \textsf{c} at range \textsf{d}. \\
    \textsf{font(d, p, v)}      & Set font property \textsf{p} $\in$ \{bold, italic, size, color, underline, name\} to \textsf{v} at \textsf{d}. \\
    \textsf{border(d, s, v)}    & Set border on side \textsf{s} $\in$ \{left, right, top, bottom, outside, all\} to style \textsf{v} at \textsf{d}. \\
    \textsf{align(d, t, v)}     & Set text property \textsf{t} $\in$ \{horizontal, vertical, orientation, wrap\} to \textsf{v} at \textsf{d}. \\
    \textsf{paste(d, s, m)} & Paste \textsf{m} $\in$ \{style, value, both\} from \textsf{s} to \textsf{d}. \\
    \textsf{autofill(d, s)}  & Auto-fill range \textsf{d} from source \textsf{s}. \\
    \bottomrule
    \end{tabularx}
    \caption{Different user actions available in spreadsheet software.}
    \label{tab:actions}
\end{table}


We reconstruct user trajectories from static, publicly available workbooks~\cite{singh2023cornet} through a three-stage pipeline: (1)~symbolic cold-start generation, (2)~LLM-assisted refinement, and (3)~human annotation. The first two stages reduce human annotation effort while introducing natural variation in construction sequences, providing annotators with an initial trajectory to work from rather than requiring them to author sequences from scratch.

\paragraph{Symbolic cold-start.}
Given a target workbook, we decompose the final state into cell-level operations and merge adjacent identical operations into range actions (e.g., \textsf{\small font(A1, bold, true)} + \textsf{\small font(A2, bold, true)} become \textsf{\small font(A1:A2, bold, true)}). To introduce variability, we sample from a space of preference settings that control merging strategy, region ordering, and formatting order.
A vision-language model annotates each sheet with semantic metadata (regions, dependencies, pasted ranges). Full details of this pipeline appear in Appendix~\ref{app:dataset-details}.

\paragraph{LLM refinement.}
An LLM-based judge-editor loop serves as a helper step, identifying and correcting obvious unnatural patterns in the symbolic sequences: consolidating scattered cell-by-cell formatting into range operations, removing stray formatting on unused areas, and adjusting sequences to follow more natural patterns.
This step reduces the annotation burden by handling mechanical corrections and all revisions are validated to ensure they reach the target workbook state.

\paragraph{Human annotation.}
Human annotators perform a final pass by correcting remaining unnatural subsequences.
As evidence of the substantial restructuring performed by annotators, the mean normalized edit distance between pre-annotation and final trajectories is 0.69 (median 0.77), and 19 of 52 trajectories were effectively rewritten from scratch (distance ${>}0.8$); see Figure~\ref{fig:edit-distance} in the Appendix.
Common refinements include grouping content with its formatting, reordering to establish structure before detail, breaking bulk inputs into individual cell entries, and trimming ranges to content boundaries.

\begin{figure}[t]
    \centering
    \includegraphics[height=3.69cm]{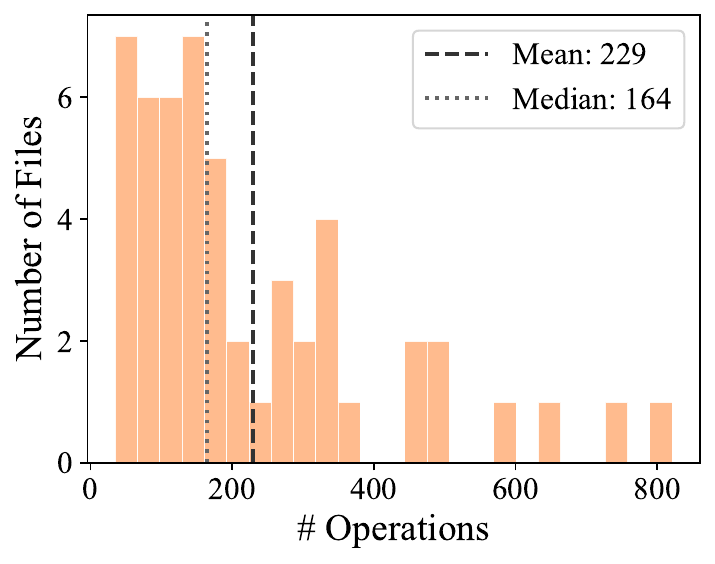}\hspace{0.02\columnwidth}%
    \includegraphics[height=3.69cm]{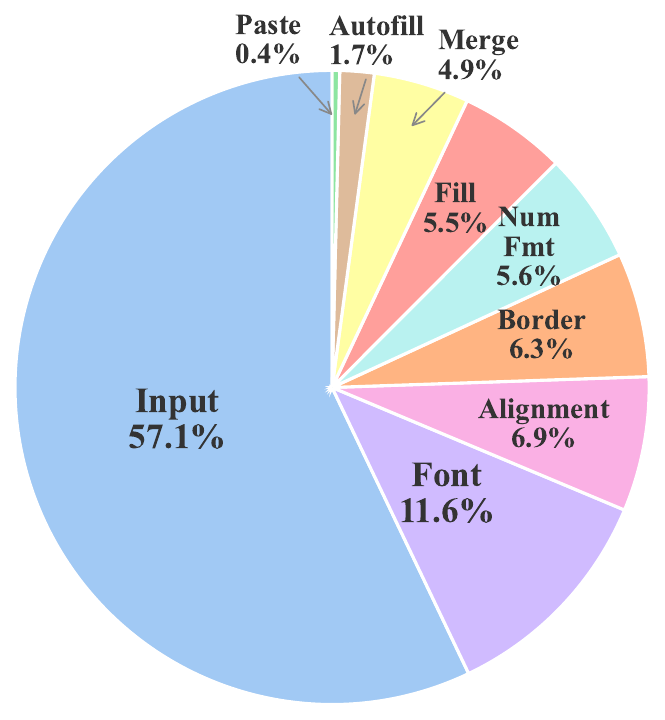}
    \caption{Dataset characteristics. (left) Distribution of file sizes by number of operations. (right) Composition of operations by category, showing diversity of formatting and content operations.}
    \label{fig:data-distribution}
\end{figure}

\paragraph{Dataset statistics.}
Our final dataset comprises 52 sheets with a combined 11,907 operations.
Sequence lengths range from 35 to 821 (mean 229, median 164).
Only one sheet is used per workbook to maximise diversity.
Figure~\ref{fig:data-distribution} shows the size distribution and operation breakdown. 

\subsection{Empirical Predictability}
\label{sec:predictability}

Not all actions in a trajectory are predictable from context alone.
To estimate an empirical upper bound on predictability, we conduct an oracle experiment: for each trajectory, we invoke frontier reasoning models at every step with the full operation history and the corresponding sheet image at that time, and ask it to predict zero or more subsequent actions.
Each model generates two generations per step, and we repeat this process across four LLMs: Opus 4.6, Opus 4.7, GPT-5.4, and GPT-5.5, all with high reasoning effort.
We then take the union of all correctly predicted (cell, property) pairs across invocations.

Across the 52 trajectories, 68\% of all ground-truth properties appear in this union, with a per-trajectory median of 66.3\% (44 of 52 trajectories exceed 50\%). This establishes that the vast majority of spreadsheet actions are, in principle, predictable from editing history, and sets a theoretical ceiling for online evaluation performance.

\begin{figure}[t]
  \centering
  \includegraphics[width=\columnwidth]{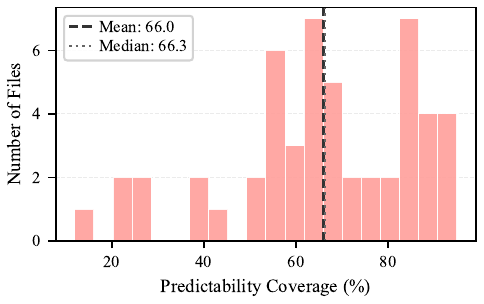}
  \caption{Distribution of predictability coverage across the 52 benchmark trajectories.}
  \label{fig:predictability-coverage}
\end{figure}

\section{Online Evaluation}
\label{sec:online_eval}

Online evaluation judges the system under the same conditions a real user operates in, where each accepted suggestion affects the future actions that a user still has to perform.
To this end, we expose a function that evaluates a given assistant using an online evaluation loop, which performs an on-policy rollout over worksheet states.
A high-level overview of this evaluation, which ignores some evaluation settings and logging metrics, is shown in Figure~\ref{alg:evaluation}.
Some acceptance criteria are explored in our experiments (Section~\ref{sec:experiments}).

The \texttt{update} step adapts the remaining future after an accepted prediction. It removes operations whose effects are already satisfied, prepends inverses for any false positives, and verifies that the adapted sequence still reaches the target state (like the partial \textsf{\small fill} in Figure~\ref{fig:introExample}).
A final \texttt{patch} step simulates the adapted future and synthesizes corrections for any residual discrepancies, ensuring the updated sequence always produces the original target state.

\begin{figure}[htb]
\begin{lstlisting}[style=pseudo,frame=single,escapeinside={(*@}{@*)}]
class ISolver:
  def predict(s: Spreadsheet,
              h: Action[]) $\rightarrow$ Action[]:
    // to implement

def evaluate(ISolver solver,
             Action[] sequence,
             bool repredict = False):
    S $\leftarrow$ Spreadsheet.empty()
    H $\leftarrow$ []
    F $\leftarrow$ sequence
    while F:
        while True:
            P $\leftarrow$ solver.predict(S, H)
            if not accept(P, H, F):
                break
            S $\leftarrow$ execute(P, S)
            F $\leftarrow$ update(F, P, S)
            H $\leftarrow$ H + P
            if not repredict: break

def update(F, P, S):
    S$_\texttt{p}$ $\leftarrow$ execute(P, S)
    S$_\texttt{f}$ $\leftarrow$ execute(F, S_p)
    F$^\prime$ $\leftarrow$ []
    // remove satisfied operations and only keep
    for op in F:
        if not satisfied(op, S$_\texttt{p}$):
            F$^\prime$ += residual(op, S$_\texttt{p}$)
    // undo false positives with inverse operations
    for p in false_pos(S$_\texttt{p}$, S$_\texttt{f}$):
        F$^\prime$ $\leftarrow$ [invert(p)] + F$^\prime$
    // verify and patch
    if execute(F$^\prime$, S$_\texttt{p}$) $\neq$ S$_\texttt{f}$:
        F$^\prime$ $\leftarrow$ patch(F$^\prime$, S$_\texttt{p}$, S$_\texttt{f}$)
    return F'
\end{lstlisting}
\caption{Pseudo-code for the interface that assistants should implement and the exposed evaluate function that performs online evaluation. Computing metrics and additional hyperparameters are omitted for brevity.}
\label{alg:evaluation}
\end{figure}

\subsection{Metrics}
\label{sec:metrics}

We define metrics at three levels of granularity.

\subsubsection{Property/Action level}

At the finest granularity, we compare individual (cell, property) pairs between the predicted state and the target state, considering only explicitly modified properties. Each pair is classified as follows. 
\textbf{True positive (TP):} the property appears in both states with the same value; the prediction correctly anticipates the user’s edit. 
\textbf{False positive (FP):} the property is present in the prediction but absent in the target, introducing an unwanted edit the user must undo. 
\textbf{False negative (FN):} the property is missing from the prediction but required by the target, leaving work the user must perform manually. 
\textbf{Mismatch (MM):} the property appears in both states but with different values, producing an incorrect edit that the user must correct.

\subsubsection{Prediction level}

A prediction refers to all actions proposed in one iteration.
These metrics can be used by \texttt{accept}. 
\textbf{Precision:} fraction of predicted properties that are correct, $\text{TP}/(\text{TP}+\text{FP}+\text{MM})$, indicating how few errors a prediction introduces. 
\textbf{User actions saved (\textsc{uas}):} the reduction in user effort if the prediction were accepted, computed by simulating acceptance and measuring how much the future diverges; positive values help the user, negative values hurt.

\subsubsection{Emulation Level}

These metrics capture system performance over a full trajectory. 
\textbf{User actions saved (\textsc{uas}):} percentage reduction in user effort, $(L_{\text{initial}} - L_{\text{final}})/L_{\text{initial}}$.  This is our primary metric for performance.
\textbf{Acceptance rate (\textsc{ar}):} fraction of predictions accepted. 
\textbf{Average precision (\textsc{prec}):} macro-average precision across all predictions. 
\textbf{Predictability Coverage (\textsc{pCov}):} fraction of \emph{predictable} (cell, property) pairs that the system actually produced, where the predictable set is the oracle union from \S\ref{sec:predictability}. This normalizes by what is learnable from context, giving a ceiling-relative measure rather than one diluted by inherently unpredictable edits.

\section{Experimental Setup}\label{sec:setup}

\subsection{Solvers}

We evaluate three families of solvers for our baselines.

\paragraph{Zero-shot LLMs.}
Predictions are made by representing actions as text, where \textsf{\small fill(A1, green)} is represented as \op{FILL | A1 | green}.
The prompt instructs the model to suggest next actions in a zero-shot manner and, besides instructions, includes: (1) the recent operation history, and (2) the valid syntax for available operations.
We evaluate GPT-5-R, GPT-5, GPT-5-R mini, and GPT-5 mini, where the ``-R'' suffix denotes models run in reasoning mode (set to low); models without the suffix are run in non-reasoning chat mode. 
Unless specified otherwise, the default model is GPT-5 (with no reasoning). 
The full prompt template is provided in the Appendix~\ref{app:prompts}. 

\paragraph{Fine-tuned SLMs.}

We fine-tune SmolLM2 models (135M and 360M parameters) on synthetically generated operation sequences. 
Training data is produced using the same symbolic pipeline described in Section~\ref{sec:benchmark}, applied to a separate pool of 12K workbooks disjoint from the evaluation set.
Each training example consists of a full operation sequence for a single sheet trajectory---averaging 163 operations---identical in format to the evaluation data.
We window each trajectory using width 32 and stride 16, yielding $\pm$45K data points for supervised fine-tuning.

\paragraph{Classical ML.}
We use four types of non-generative predictors on operation sequences: trained $n$-gram, online $n$-gram, LSTM, and XGBoost.
The trained $n$-gram, LSTM, and XGBoost models are trained on the same synthetic data as the SLMs.
The online $n$-gram requires no training and learns patterns from past actions within each trajectory.
Each predicts the next operation from a sliding window of recent actions using relative features (offsets from the previous action) rather than absolute features (cell addresses). 
We found this to generalize better across trajectories in preliminary experiments.
Full architecture details are in Appendix~\ref{app:classical-solvers}.

\subsection{Evaluation Settings}

We use two evaluation settings that differ in how many actions the solver produces per invocation.

\begin{figure}[t]
  \centering
  \includegraphics[width=\columnwidth]{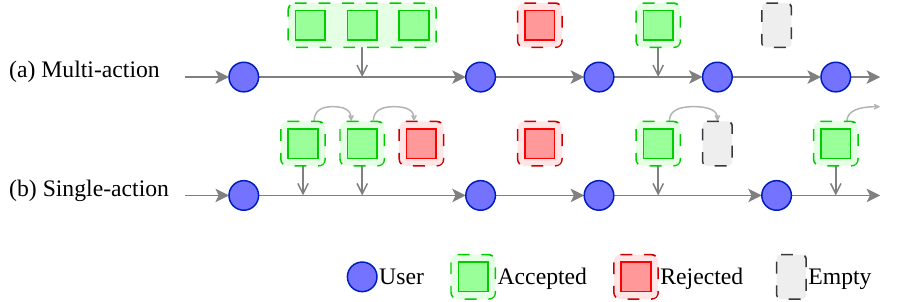}
  \caption{%
    Two prediction settings. (a)~\emph{Single-action} (k\,$=$\,1, repredict):
    the solver is queried for one op at a time; after each acceptance the
    solver is re-queried before the next user step. (b)~\emph{Multi-action}
    (k\,$\geq$\,1): the solver may emit a block of up to $k$ ops per query,
    and the cursor advances by the length of any accepted block.%
  }
  \label{fig:predict-modes}
\end{figure}

\paragraph{Multi-action prediction.}
The solver is invoked once on the current history and predicts a variable-length sequence of actions in a single shot.
This setting tests both prediction quality and the solver's ability to decide when to stop: longer correct predictions save more user actions, but stopping too late introduces false positives that the user must reject or undo.
This is a natural setting for zero-shot LLMs, which can produce arbitrary-length completions and decide internally when to stop.

\paragraph{Single-action repredict.}
The solver predicts exactly one action per invocation.
If accepted, it is immediately re-invoked on the updated history to predict the subsequent action; this continues until a prediction is rejected.
We use this setting to evaluate solvers that (for now) lack the capacity to autonomously decide when to stop predicting, such as fine-tuned SLMs and classical sequence models.
It removes the stopping burden entirely: the evaluation loop controls termination via the acceptance heuristic.
This is analogous to a user accepting word-by-word completions (repeatedly pressing \emph{tab} in a code editor).

\subsection{Hyperparameters}

We vary four key hyperparameters that control the evaluation.
\textbf{Prediction stride} ($s$): the number of user actions between successive solver invocations; $s{=}1$ invokes the solver after every action.
\textbf{Context window} ($c$): the number of recent operations provided to the solver as history.
\textbf{Max actions per prediction} ($m$): the number of actions the solver may produce per invocation; $m{=}\infty$ for multi-action, $m=1$ for single-action re-predict.
\textbf{Context shortening}: whether long cell values in the history are truncated to a fixed character limit before being sent to the solver, reducing token usage.
Unless stated otherwise, we use $s=1$, $c=32$, $m=\infty$ (multi-action) or $m=1$ (single-action re-predict), and context shortening enabled.

\subsection{Acceptance Heuristics}

The acceptance heuristic (\texttt{accept} in Figure~\ref{alg:evaluation}) determines whether a prediction is applied or rejected.
We experiment with several heuristics that combine precision and savings constraints, as shown in Table~\ref{tab:heuristics-results}.


The \textsc{greedy} heuristic models a user who accepts any suggestion that advances their goal, regardless of minor errors.
Precision-only heuristics (\textsc{p90}, \textsc{p60}) model a user who accepts predictions that look mostly correct, even if they do not reduce overall effort.
The \textsc{always} baseline accepts every prediction regardless of quality, serving as an upper bound on acceptance rate and a stress test for the ground-truth adaptation procedure.

\section{Experiments}\label{sec:experiments}


\begin{table}[t]
  \caption{Model comparison in single-action repredict ($s{=}1$, $c{=}32$, \textsc{greedy}). -R means reasoning (low).}
  \label{tab:models-repredict}
  \centering
  \small
  \begin{tabular}{lrrrr}
    \toprule
    \textsc{model} & \textsc{uas}$\uparrow$ & \textsc{ar}$\uparrow$ & \textsc{prec}$\uparrow$ & \textsc{pCov}$\uparrow$ \\
    \midrule
    \multicolumn{5}{l}{\itshape Zero-shot LLMs} \\
    GPT-5-R      & 32.7 & 29.4 & 41.6 & 24.8 \\
    GPT-5-R mini & 28.2 & 25.5 & 37.0 & 20.9 \\
    GPT-5         & 27.4 & 30.9 & 44.8 & 20.7 \\
    GPT-5 mini    & 18.0 & 16.8 & 21.9 & 10.7 \\
    \midrule
    \multicolumn{5}{l}{\itshape SLMs (base)} \\
    SmolLM2-360M         & 21.7 & 22.3 & 29.7 &  9.6 \\
    SmolLM2-135M         & 18.3 & 19.0 & 24.8 &  8.9 \\
    \multicolumn{5}{l}{\itshape SLMs (finetuned)} \\
    FT-SmolLM2-360M      & 26.8 & 26.8 & 33.7 & 13.7 \\
    FT-SmolLM2-135M      & 23.2 & 23.1 & 30.6 & 13.0 \\
    \midrule
    \multicolumn{5}{l}{\itshape Classical sequence models} \\
    LSTM                 &  5.7 &  5.5 & 12.4 &  2.4 \\
    Trained $n$-gram       &  3.8 &  3.9 & 11.9 &  0.7 \\
    XGBoost              &  2.9 &  2.3 &  6.5 &  1.0 \\
    Online $n$-gram        & 12.0 & 14.7 & 20.4 & 11.1 \\
    \bottomrule
  \end{tabular}
\end{table}

\begin{table}[t]
  \caption{Hyperparameter ablation (single-action repredict, GPT-5, \textsc{greedy}). Defaults marked $^{*}$.}
  \label{tab:ablation-single}
  \centering
  \small
  \begin{tabular}{@{}llrrrr@{}}
    \toprule
    \textsc{param} & \textsc{value} & \textsc{uas}$\uparrow$ & \textsc{ar}$\uparrow$ & \textsc{prec}$\uparrow$ & \textsc{pCov}$\uparrow$ \\
    \midrule
    \multirow{4}{*}{Stride ($s$)}
     & 1$^{*}$  & 27.4 & 30.9 & 44.8 & 20.7 \\
     & 2        & 22.6 & 36.5 & 48.4 & 15.9 \\
     & 4        & 16.8 & 42.3 & 53.2 &  9.4 \\
     & 8        & 10.6 & 43.7 & 55.1 &  7.1 \\
    \midrule
    \multirow{4}{*}{Context ($c$)}
     & 8        & 19.9 & 24.1 & 39.7 & 13.6 \\
     & 32$^{*}$ & 27.4 & 30.9 & 44.8 & 20.7 \\
     & 128      & 30.0 & 32.5 & 47.8 & 28.5 \\
     & 512      & 30.8 & 33.7 & 47.9 & 31.0 \\
    \midrule
    \multirow{2}{*}{Shortening}
     & on$^{*}$ & 27.4 & 30.9 & 44.8 & 20.7 \\
     & off      & 27.7 & 31.2 & 44.3 & 21.4 \\
    \midrule
    \multirow{2}{*}{Repredict}
     & on$^{*}$ & 27.4 & 30.9 & 44.8 & 20.7 \\
     & off      & 20.3 & 30.6 & 44.1 & 15.2 \\
    \bottomrule
  \end{tabular}
\end{table}

\begin{figure*}[t]
    \centering
    \includegraphics[width=\textwidth]{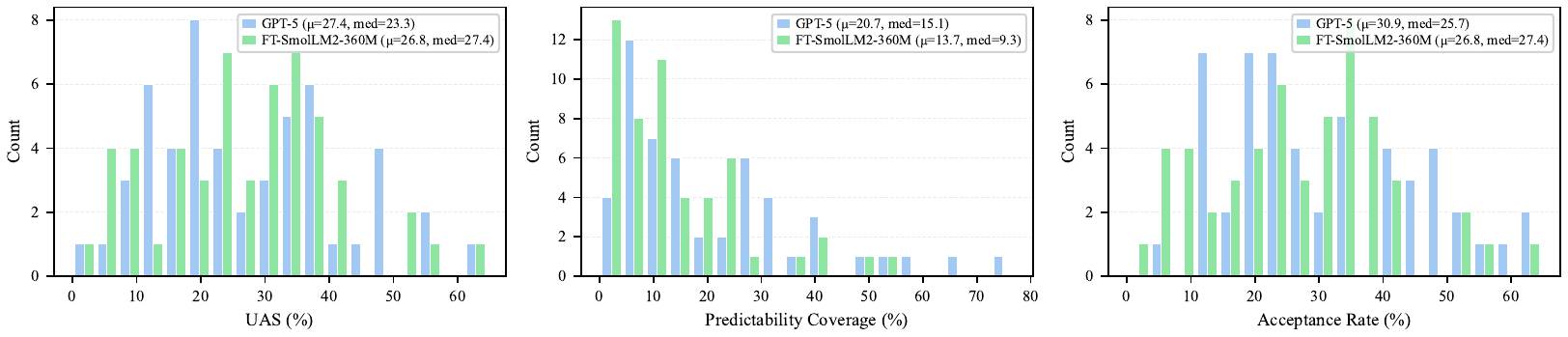}
    \caption{Per-file distributions across the 52 benchmark trajectories, comparing GPT-5 (blue) and FT-SmolLM2-360M (green), both run with single-action repredict and default settings. (Left) User Actions Saved. (Center) Predictability Coverage, the fraction of properties an oracle deemed predictable that the model correctly produced. (Right) Acceptance Rate. Legend entries show mean ($\mu$) and median for each model.}
    \label{fig:file-distributions}
\end{figure*}

\begin{figure*}[t]
    \centering
    \includegraphics[width=\textwidth]{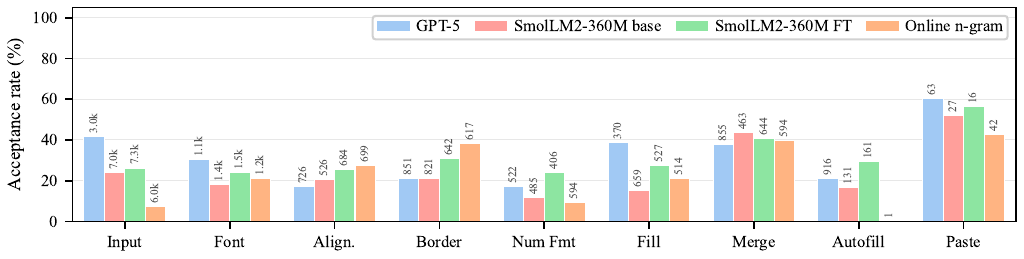}
    \caption{Per-category acceptance rate across four solvers. Numbers above each bar give the count of predictions that solver made in that category.}
    \label{fig:op-category-acceptance}
\end{figure*}

\begin{table}[t]
  \caption{Multi-action prediction ($s{=}1$, $c{=}32$, \textsc{greedy}).}
  \label{tab:models-multi}
  \centering
  \small
  \begin{tabular}{lrrrr}
    \toprule
    \textsc{model} & \textsc{uas}$\uparrow$ & \textsc{ar}$\uparrow$ & \textsc{prec}$\uparrow$ & \textsc{pCov}$\uparrow$ \\
    \midrule
    GPT-5-R  & 26.6 & 13.3 & 27.8 & 30.7 \\
    GPT-5       & 22.3 & 20.0 & 36.0 & 19.6 \\
    GPT-5-R mini       & 21.3 &  9.6 & 23.2 & 24.3 \\
    GPT-5 mini  & 20.1 & 10.1 & 19.0 & 17.3 \\
    \bottomrule
  \end{tabular}
\end{table}

\begin{table}[t]
  \caption{Effect of acceptance heuristics (multi-action, GPT-5, $s{=}1$, $c{=}32$). \textit{Rule}: $p$ = precision, $\ell$ = steps saved.}
  \label{tab:heuristics-results}
  \centering
  \small
  \setlength{\tabcolsep}{4pt}
  \begin{tabular}{@{}llrrrr@{}}
    \toprule
    \textsc{heur.} & \textsc{rule} & \textsc{uas}$\uparrow$ & \textsc{ar}$\uparrow$ & \textsc{prec}$\uparrow$ & \textsc{pCov}$\uparrow$ \\
    \midrule
    \textsc{greedy}   & $\ell{\geq}1$                & 22.3 & 20.0 & 36.0 & 19.6 \\
    \textsc{hybrid-1} & $p{\geq}.9\,,\,\ell{\geq}1$  & 21.8 & 16.6 & 39.4 & 17.3 \\
    \textsc{greedy-2} & $\ell{\geq}2$                & 20.3 & 10.5 & 37.6 & 15.9 \\
    \textsc{hybrid-2} & $p{=}1\,,\,\ell{\geq}2$      & 17.5 &  7.9 & 40.1 & 10.7 \\
    \textsc{p100}     & $p{=}1$                      & 19.9 & 21.8 & 38.2 & 20.1 \\
    \textsc{p90}      & $p{\geq}.9$                  & 17.0 & 23.3 & 36.0 & 25.5 \\
    \textsc{p60}      & $p{\geq}.6$                  & 13.3 & 26.9 & 32.9 & 33.0 \\
    \textsc{always}   & ---                          & $-$19.2 & 100.0 & 9.3 & 8.1 \\
    \bottomrule
  \end{tabular}
\end{table}

\begin{table}[t]
  \caption{Hyperparameter ablation (multi-action, GPT-5, \textsc{greedy}). Defaults marked $^{*}$.}
  \label{tab:ablation-multi}
  \centering
  \small
  \begin{tabular}{@{}llrrrr@{}}
    \toprule
    \textsc{param} & \textsc{value} & \textsc{uas}$\uparrow$ & \textsc{ar}$\uparrow$ & \textsc{prec}$\uparrow$ & \textsc{pCov}$\uparrow$ \\
    \midrule
    \multirow{4}{*}{Stride ($s$)}
     & 1$^{*}$    & 22.3 & 20.0 & 36.0 & 19.6 \\
     & 2          & 19.5 & 24.4 & 39.4 & 16.0 \\
     & 4          & 14.7 & 33.4 & 45.1 & 12.2 \\
     & 8          &  9.8 & 36.5 & 49.4 &  6.8 \\
    \midrule
    \multirow{5}{*}{Context ($c$)}
     & 8          & 16.2 & 17.5 & 33.8 & 12.3 \\
     & 32$^{*}$   & 22.3 & 20.0 & 36.0 & 19.6 \\
     & 128        & 27.6 & 19.7 & 37.3 & 30.0 \\
     & 512        & 26.2 & 19.0 & 35.9 & 32.0 \\
     & 2048       & 27.4 & 19.4 & 36.6 & 33.0 \\
    \midrule
    \multirow{2}{*}{Shortening}
     & on$^{*}$   & 22.3 & 20.0 & 36.0 & 19.6 \\
     & off        & 24.2 & 21.5 & 38.4 & 21.4 \\
    \midrule
    \multirow{5}{*}{Num ops ($m$)}
     & 1          & 20.3 & 30.6 & 44.1 & 15.2 \\
     & 4          & 20.8 & 23.1 & 39.2 & 19.4 \\
     & 8          & 22.1 & 19.4 & 36.1 & 18.9 \\
     & 16         & 21.8 & 20.6 & 36.2 & 20.2 \\
     & $\infty^{*}$ & 22.3 & 20.0 & 36.0 & 19.6 \\
    \bottomrule
  \end{tabular}
\end{table}

\begin{figure}[t]
    \centering
    \includegraphics[width=0.85\columnwidth]{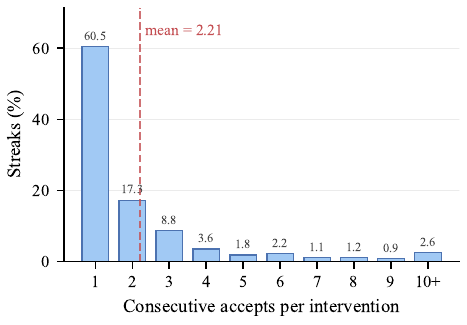}
    \caption{Distribution of acceptance streak lengths for GPT-5 in the single-action repredict setting (1{,}204 streaks across 52 trajectories). Most streaks are short, but a heavy tail of longer runs (3+ consecutive accepts) drives most of the realised user-action savings.}
    \label{fig:acceptance-streaks}
\end{figure}

\begin{figure}[t]
    \centering
    \includegraphics[width=0.9\columnwidth]{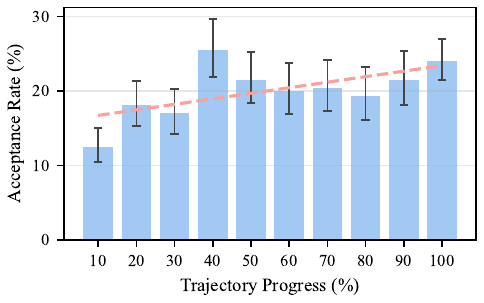}
    \caption{Acceptance rate by trajectory progress (10\% bins). As users progress through their spreadsheet editing task, acceptance rates increase, suggesting predictions become more accurate as more context becomes available. (uses default settings)}
    \label{fig:acceptance-trajectory-progress}
\end{figure}

\begin{figure}[t]
    \centering
    \includegraphics[width=0.9\columnwidth]{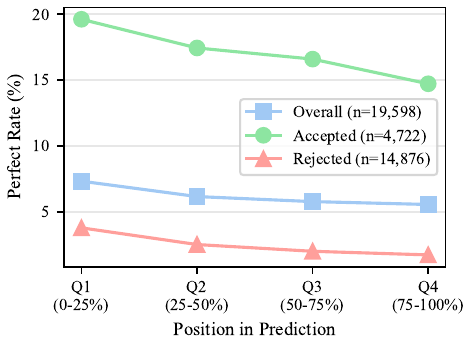}
    \caption{Perfect rate (precision = 100\%) by operation position within predictions (uses default settings). Operations are grouped by their normalized position (quartile) within each prediction. Accepted predictions show consistently higher perfect rates across all positions compared to rejected predictions.}
    \label{fig:perfect-by-quartile}
\end{figure}

\paragraph{The task is learnable.}

We observe a correlation between model capability and prediction performance (Tables~\ref{tab:models-multi} and~\ref{tab:models-repredict}).
The consistent gradient from GPT-5 mini (20.1\% \textsc{uas} multi-action, 18.0\% single-action) to GPT-5-R (26.6\% multi-action, 32.7\% single-action) confirms that action prediction is a learnable task, not lucky guessing.
Stronger models systematically perform better, suggesting significant room for improvement through specialized training or architectures optimized for this domain.
Fine-tuning further supports this: SmolLM2-360M improves from 21.7\% to 26.8\% \textsc{uas} and SmolLM2-135M from 18.3\% to 23.2\% after fine-tuning on operation sequences (Table~\ref{tab:models-repredict}), narrowing the gap to GPT-5 (27.4\%) despite their much smaller size.

\paragraph{Abstention is key.}

The \textsc{always} heuristic, which accepts every non-empty prediction, yields strongly negative savings ($-19.2$\% \textsc{uas}) at only 9.3\% precision (Table~\ref{tab:heuristics-results}); in fact, 51 of 52 trajectories hit the enforced user-step cap of 120\% of ground truth, so the unconstrained number would be far worse.
This illustrates that LLMs do not natively know when to abstain (no intent is available) or when to stop predicting, two key areas to focus on for future predictors.
Since smaller models and classical sequence solvers cannot be fairly evaluated in the multi-action setting, we additionally report results in the single-action repredict setting (Table~\ref{tab:models-repredict}), an equivalent setup that levels the playing field by restricting every solver to one action per call.
Under this protocol, \mbox{SmolLM2-360M FT} (26.8\% \textsc{uas}) approaches GPT-5 (27.4\%) and even the training-free \textsc{online $n$-gram} (12.0\%) becomes a non-trivial baseline, highlighting the potential of smaller models and motivating research on explicit abstention objectives rather than on raw generative capacity alone.

\paragraph{Cheaper predictions or triggers.}

The stride $s$ dramatically affects system behavior (Table~\ref{tab:ablation-multi}).
At $s{=}1$, the system attempts a prediction after every user action, achieving the highest \textsc{uas} (22.3\%) despite the lowest acceptance rate (20.0\%).
At $s{=}8$, the acceptance rate climbs to 36.5\% but \textsc{uas} collapse to 9.8\%.
The same pattern holds in the single-action setting (Table~\ref{tab:ablation-single}): $s{=}1\!\to\!8$ moves \textsc{ar} from 30.9\% to 43.7\% while \textsc{uas} falls from 27.4\% to 10.6\%.
A cheap trigger that can be invoked at every iteration is therefore a key focus area to improve overall utility.

\paragraph{Easier and harder actions.}
Per-category acceptance rates (Figure~\ref{fig:op-category-acceptance}) split cleanly into two groups across all solvers: content-heavy operations (\op{input}, \op{paste}, \op{fill}, \op{merge}) are accepted at noticeably higher rates than presentational ones (\op{align}, \op{number format}, \op{border}).
GPT-5 dominates on \op{input}, reflecting its strong language prior on cell content, while the training-free \op{online $n$-gram} holds up well on structural categories driven by local repetition (\op{merge}, \op{border}, \op{paste}) but collapses on semantic categories such as \op{input}.

Comparing the base and fine-tuned SmolLM2-360M makes the effect of in-domain training concrete: the largest accuracy gains land on the categories the base handled worst---number format, \op{fill}, \op{border}, and \op{autofill} all jump by roughly 10 points---while content categories shift only marginally.
At the same time, the fine-tuned model issues fewer \op{border} and \op{fill} predictions and more \op{merge}, \op{align}, and \op{autofill} ones, suggesting it learns both \emph{when} to predict each category and \emph{how} to predict it more reliably.

\paragraph{Acceptance streaks.}
We define an \emph{acceptance streak} as a maximal run of consecutive accepted single-action predictions between two user interventions (Figure~\ref{fig:acceptance-streaks}). The distribution for our base run is heavy-tailed: most streaks are short (median $1$), yet the mean of $2.21$ implies that each \emph{successful} intervention commits roughly twice as many operations as the user would otherwise type, and $5\%$ of streaks chain seven or more accepts. The other solvers we evaluated in Table~\ref{tab:models-repredict} exhibit the same qualitative pattern, with mean streak lengths clustered in the $1.6$--$2.3$ range. This burst-like behaviour reconciles the seemingly modest per-prediction acceptance rate ($\sim$32\%) with the substantially larger reported \textsc{uas}: predictors do not save effort uniformly, but rather lock onto locally repetitive structure (row fills, border propagation, repeated formatting) and chain several confident predictions before the user must reassert control.

\paragraph{Conservative acceptance is not strictly better.}

The \textsc{greedy-2} heuristic demands higher utility than \textsc{greedy}, and \textsc{hybrid-2} layers an additional perfect-precision constraint on top (Table~\ref{tab:heuristics-results}).
As expected, acceptance rates fall (20.0\% $\to$ 10.5\% $\to$ 7.9\%), but \textsc{uas} also falls (22.3\% $\to$ 20.3\% $\to$ 17.5\%).
Opportunistic acceptance (\textsc{greedy}) yields the best cumulative benefit, while strict filters discard borderline-but-useful predictions faster than they remove harmful ones.

\paragraph{Precision alone does not ensure utility.}

Even at \textsc{p90}, which accepts any prediction with $\geq 90\%$ precision regardless of utility, the system delivers only 17.0\% \textsc{uas} -- 4.8 points below \textsc{hybrid-1} (21.8\%) and 5.3 points below \textsc{greedy} (22.3\%) (Table~\ref{tab:heuristics-results}).
\textsc{p100} (perfect precision) does better at 19.9\% but is still surpassed by \textsc{hybrid-1}.
Precision measures correctness of \emph{individual} predicted properties, not whether a prediction actually advances the user toward completion; a heuristic that ignores utility leaves savings on the table.

\paragraph{The right context length.}

Context length has a clear effect on performance, but with quickly diminishing returns (Tables~\ref{tab:ablation-multi} and~\ref{tab:ablation-single}).
In the multi-action setting, increasing the window from 32 to 128 prior operations lifts \textsc{uas} from 22.3\% to 27.6\%, while leaving the acceptance rate essentially unchanged.
Gains taper rapidly beyond this point: 512 and 2048 contexts yield 26.2\% and 27.4\% \textsc{uas} respectively, indistinguishable from 128.
The single-action setting shows the same saturation (27.4\% at $c{=}32$, 30.0\% at 128, 30.8\% at 512).
This suggests that while long-range history helps, most of the predictive signal resides within the most recent $\sim$128 operations.


\paragraph{The right prediction length.}
Unlimited prediction scope performs best in the multi-action setting; constraining the per-call budget never improves \textsc{uas} (Table~\ref{tab:ablation-multi}).
Capping the model at $m{=}1$ inflates the acceptance rate to 30.6\% but yields only 20.3\% \textsc{uas}, because each accepted prediction now saves at most one step.
Mid-range caps ($m{=}4$, $m{=}8$, $m{=}16$) all hover around 21--22\% \textsc{uas}, roughly growing lightly with the limit.
This suggests the model already self-regulates prediction length based on confidence, and external limits prevent it from exploiting high‑confidence opportunities. Notably, the $m=16$ setting performs worse than $m=8$ on \textsc{uas}, this could be due to higher caps encouraging the model to “fill” its budget with lower-confidence edits rather than stopping naturally.

\paragraph{Acceptance rate increases as trajectories progress.}

The acceptance rate rises with trajectory progress, from $\sim$12.5\% in the first 10\% of a trajectory to $\sim$24\% by the end (Figure~\ref{fig:acceptance-trajectory-progress}).
This indicates a cold-start problem: without established patterns to recognize, early predictions are largely guesses.
As context grows, the model identifies repetitive formatting, table structures, and user-specific conventions that make later predictions more reliable.
This suggests that adaptive triggering---predicting less frequently early in trajectories and more frequently later---could improve overall efficiency by avoiding the high rejection rate of early predictions.

\paragraph{Overpredictions.}

Figure~\ref{fig:perfect-by-quartile} shows how the perfect-precision rate of an action varies with its position inside the prediction.
For \emph{accepted} predictions, the perfect rate decays monotonically from $\sim$19.5\% in the first quartile to $\sim$14.7\% in the last: when the assistant knows what to do, it starts by doing the right thing and then drifts.
\emph{Rejected} predictions are subpar throughout and follow the opposite, slightly declining trend ($\sim$3.5\% $\to$ $\sim$1.5\%).
This suggests that abstention is relevant at two granularities: at the prediction level (to suppress entirely-bad predictions) and at the action level within an otherwise good prediction (to truncate before the tail degrades).
The latter is reinforced by our observation that a substantial share of non-empty predictions contribute at least one true positive to the final state yet are still rejected---a gap that represents real improvement potential, since many predictions contain useful content but are discarded because of accompanying errors that violate acceptance heuristics.

\section{Related Work}

\paragraph{Spreadsheet automation.}
Prior work on spreadsheet automation spans two paradigms.
\emph{Natural-language-to-action} systems translate explicit user commands into operations: SheetCopilot~\citep{li2023sheetcopilot} and SpreadsheetBench~\citep{ma2024spreadsheetbench} benchmark this approach, while TableTalk~\citep{liang2025tabletalk} studies how users interact with such assistants and finds that for routine edits, the overhead of formulating prompts exceeds the cost of direct manipulation.
\emph{Action-to-action} systems instead suggest completions from prior user actions without explicit invocation: FlashFill~\citep{gulwani2011automating} synthesises string transformations from input-output examples, FlashRelate~\citep{barowy2015flashrelate} extracts relational data from semi-structured spreadsheets, and SpreadsheetCoder~\citep{chen2021spreadsheetcoder} predicts formulas from tabular context.
Our benchmark targets a third, underexplored category: \emph{modeless} action suggestion, where the system continuously observes user edits and proactively suggests next actions without being invoked.

\paragraph{Modeless edit suggestion and code completion.}
Modeless suggestion systems, which surface predictions without explicit user invocation, are well established in code editing.
Early symbolic systems like Blue-Pencil~\citep{miltner2019fly} suggested repetitive edits, IntelliCode~\citep{svyatkovskiy2020intellicode} extended this to full-line completion, and modern assistants like GitHub Copilot can suggest multi-line blocks~\citep{mastropaolo2023robustness}.
\citet{izadi2024language} provide a practical evaluation of language models for code completion, establishing methodologies that inform our evaluation design.
\citet{mozannar2024show} study when to surface a suggestion in AI-assisted programming, directly motivating our investigation of prediction stride and acceptance heuristics.
In GUI automation, ASSISTGUI \citep{gao2024assistgui} benchmarks task-oriented desktop automation, a setting that shares the sequential, state-modifying nature of spreadsheet editing.
Most directly related, \citet{shaikh2026learning} learn next-action predictors from human-computer interaction traces in desktop UI settings; our work addresses the analogous problem in spreadsheets.

\paragraph{Interactive evaluation.}
Evaluating interactive assistants under realistic conditions is a cross-domain challenge.
In information retrieval, \citet{borlund2009user} argues for user-centred evaluation that assesses systems as they would actually be used.
In recommender systems, \citet{castells2022offline} show that offline metrics diverge from actual user utility due to the unobserved-truth problem, and \citet{ferrari2022offline} demonstrate that offline evaluation of interacting recommendations fails to capture their combined impact.
Offline policy evaluation in sequential settings~\citep{mandel2014offline} faces analogous distribution shift.
Spreadsheet assistance exhibits similar issues: predictions reshape state, users vary in which suggestions they accept, and the same final workbook can result from many different action sequences.
Our online evaluation loop addresses these challenges by simulating acceptance decisions and dynamically adapting the ground truth.

\section{Conclusion}

We introduce the first benchmark for the next action prediction task in spreadsheets.
Through synthetic generation and refinement, we create 52 benchmark tasks that span 12K actions in total.
Online evaluation allows us to evaluate how users perceive using the predictor.
Our experiments demonstrate that this benchmark can be used to gain actionable insights for predictors, such as room for improvement by abstention, cheap triggers, stopping criteria, and more.

\section*{Impact Statement}

This paper presents work whose goal is to advance the field of Machine Learning. There are many potential societal consequences of our work, none which we feel must be specifically highlighted here.
Note that, whereas LLMs are used in our baselines to demonstrate the benchmark, we specifically encourage research on less energy intensive methods to solve this problem.





\bibliography{example_paper}
\bibliographystyle{icml2026}

\newpage
\appendix
\onecolumn

\section{Dataset Construction Details}\label{app:dataset-details}
We leverage a combination of symbolic heuristics, LLM refinement and human refinement to reconstruct user trajectories from static workbooks.
First, we leverage an LLM to annotate spreadsheets with semantic metadata, such as table and header information.
Second, we use symbolic heuristics to generate low-level steps, like inserting values or applying formatting.
Third, an LLM is used to make the heuristic sequences more natural by reordering and combining actions.
Fourth, we perform a manual inspection over these trajectories and resolve last-mile inconsistencies to ensure high quality data.
The following four subsections describe these in detail. 

\subsection{Worksheet Annotation}

Users do not create entire spreadsheets value by value.
Data is often entered from external sources (other workbook or web pages) and repetitive parts are created by copy and pasting a related part of the sheet and then adapting it.
As these relations are hard to determine symbolically, we use a vision-language model (VLM) to annotate each spreadsheet with four types of metadata.

\begin{enumerate}
    \item \textbf{Regions} are distinct areas of the spreadsheet that have some logical meaning, such as tables (including surrounding headers and labels), title blocks, or calculation areas.
    We imagine that a user typically finishes each region before moving to the next. For each region, the VLM predicts:
    \begin{itemize}
        \item A \emph{type} classification (e.g., data tables, sheet headers, text descriptions, formula calculations).
        \item Optional \emph{closing operations}---formatting or data-entry operations that a user would naturally perform after the main structure is built (e.g., applying a highlight color to data cells after the table skeleton is complete, or bolding headers as a final touch).
    \end{itemize}
    
    \item \textbf{Region dependencies} are temporal dependencies between regions.
    Some dependencies can be inferred symbolically from formula references, but others are semantic: for example, a summary table should be built after the data it summarizes, even if there is no direct formula link.
    
    \item \textbf{Pasted ranges} are contiguous areas where values were likely pasted in bulk from an external source (CSV, clipboard, or another file), as opposed to being entered cell by cell.
    Cells containing formulas are excluded, as formulas are typically entered individually.
    
    \item \textbf{Similarly formatted regions} are groups of two or more regions that share nearly identical formatting, suggesting that the user copied formatting or duplicated a template.
    For each group, a \emph{format paste type} is also predicted: \emph{format} (only formatting style was copied and the values were ignored) or \emph{full} (the region was fully cloned, then values were overwritten).
\end{enumerate}

The input to the VLM consists of (1) a screenshot of the sheet, (2) information about contiguous regions of cells, (3) information about cells with formulas, (4) merged cells.
It is instructed to assign each cell to at least one region, and is iteratively instructed to refine its output as long as that constraint is not valid.



\subsection{Heuristic trajectories}\label{ssec:trajectories}

Given the annotated metadata, we then heuristically build a first sequence of steps.
For each individual cell and the annotated pasted regions, it is now straight-forward to symbolically generate a sequence of actions that re-generates the spreadsheet.
This would look very unnatural, however, because applying the same formatting to contiguous ranges can be done in a single action and---except for some hard constraints like first entering a value and then applying formatting---different users likely create spreadsheets in different orders.

To get more diversity, we first sample a combination of preference settings.
These settings control how operations are merged, how regions are ordered, and how operations are ordered within regions (if applicable).
Table~\ref{tab:symbolicparameters} shows a list of all such settings and the weights for their options.

First, cell operations are grouped into range operations.
Identical operations on contiguous ranges are merged into a single action (\textsf{\small font(A1, bold, true) + font(A2, bold, true)} $\rightarrow$ \textsf{\small font(A1:A2, bold, true)}).
Border operations are merged whenever the resulting action can be applied with a single button (iteratively, greedily searched for starting from the first border property that was not merged).
Pasted regions are merged based on the predicted metadata.

Second, the operations are ordered across and within regions (depending on settings).
Regions without predicted (or explicit) dependencies are ordered according to the tie breaker.
If the format order is deterministic, it is still sampled for new spreadsheets, but choices are simply remembered.
This means that one spreadsheet can have font colors always set before font properties, another spreadsheet can have the converse, and another spreadsheet can have a different order for each cell.

\begin{table*}[t]
    \centering
    \small
    \begin{minipage}[t]{0.49\textwidth}
    \centering
    \begin{tabularx}{\textwidth}{lX}
    \toprule
    \textbf{Parameter} & \textbf{Values (weight)} \\
    \midrule
    \multicolumn{2}{c}{\bfseries Operation merging} \\
    \midrule
    Merge scope                  & global (.75), per-region (.25) \\
    Merge order                  & row-first (.60), col-first (.40) \\
    Merge input operations       & true (.95), false (.05) \\
    $\quad\hookrightarrow$ group by type        & true (.80), false (.20) \\
    $\quad\hookrightarrow$ data tables only$^\dag$ & true (.75), false (.25) \\
    Merge pasted ranges$^\dag$   & true (.75), false (.25) \\
    Detect paste pattern$^\dag$  & true (.50), false (.50) \\
    $\quad\hookrightarrow$ Paste type$^\dag$    & auto (.70), format (.20), full (.10) \\
    $\quad\hookrightarrow$ Paste batching       & atomic (.50), sequential (.50) \\
    \bottomrule
    \end{tabularx}
    \end{minipage}\hfill
    \begin{minipage}[t]{0.49\textwidth}
    \centering
    \begin{tabularx}{\textwidth}{lX}
    \toprule
    \textbf{Parameter} & \textbf{Values (weight)} \\
    \midrule
    \multicolumn{2}{c}{\bfseries Ordering} \\
    \midrule
    Region ordering$^\dag$       & metadata (.75), none (.20), tie (.05) \\
    $\quad\hookrightarrow$ Tie breaker$^\dag$   & row (.60), col (.35), metadata (.05) \\
    Cross-region sorting         & enabled (.60), disabled (.40) \\
    Cell sort order              & row (.60), col (.40) \\
    Format order                 & random (.75), deterministic (.25) \\
    Use closing operations$^\dag$ & true (.75), false (.25) \\
    \bottomrule
    \end{tabularx}
    \end{minipage}
    \caption{Parameters for symbolic sequence generation. Parameters marked
    with $\hookrightarrow$ depend on the parent parameter; those marked
    $^\dag$ rely on predicted metadata. Weights were chosen based on
    intuition.}
    \label{tab:symbolicparameters}
\end{table*}

\subsection{LLM Refinement}

The symbolic step generator produces sequences that are functionally correct, but often read too rigid and mechanical, and may not reflect how a human would naturally build the sheet.
This stage refines such sequences using an LLM-based judge-editor loop with validation.

At the start of each iteration, a judge model evaluates whether the original sequence appears human-like and, if not, provides feedback.
The input to the judge is a screenshot of the sheet, the current sequence of actions, and in later interactions an overview of the changes that have already been made.
The judge is asked to pay attention to dependencies, input granularity, formatting, merging and layout, repetitive patterns, default values, borders, pasting patterns.
It is instructed to include natural variations (not always exactly row or column first) and correct user mistakes (to not penalize correct predictions during evaluation).
The feedback is not structured, as it is parsed by a subsequent LLM call.%

If any feedback is provided, a separate editor model is asked to rewrite the sequence of actions.
The input to the editor is the judge feedback and the current sequence of actions.
The output is a revised sequence of actions.
This new sequence is validated by executing the operations and verifying that the resulting state matches the target (modulo some allowed noise, such as operations that have no effect).
Editing is retried if it does not match up to 2 times, after which the last judging is also retried.
This loop continues until the judge accepts the sequence or a maximum number of iterations is reached (set to 4).

The following are examples of refinements that we have observed from LLMs. 
(1) Scattered cell-by-cell formatting: consolidate fragmented border or format operations on adjacent cells into a single range operation on the actual content area.
(2) Stray formatting on unused areas: remove formatting operations applied to empty cells or columns outside the content area.
(3) Number format on text cells: remove number or date formatting applied to cells containing text.

\subsection{Human Refinement} 

While these are good first steps toward scaling up generation, heuristics and prompt instructions do not always generalize well to all potential spreadsheets.
These steps can occasionally produce sequences that, while valid, do not seem natural.
Human annotators (the authors) perform a final pass by watching the spreadsheet being built action-by-action and spotting unnatural subsequences of actions.

The following are examples of refinements that were performed by the human annotators.
(1) Grouping content with formatting: move formatting operations closer to the content they style (e.g., green/red cell-fill actions should sit next to their true/false values) rather than being scattered.
(2) Reordering to establish structure before detail: move operations for certain areas (like table header rows and columns) earlier in the sequence before filling in the detailed content within those sections.
(3) Breaking down bulk input actions to cell-by-cell entry: split certain large input operations that look hand-typed values into individual cell actions.
(4) Trimming ranges to content boundaries: replace input ranges that include trailing empty cells with ranges covering only actual values, avoiding implicit empty-cell padding.

\begin{figure}[t]
  \centering
  \includegraphics[width=\columnwidth]{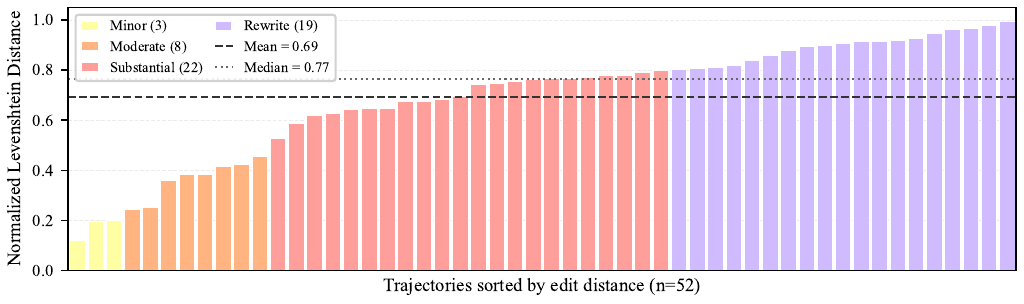}
  \caption{%
    Per-trajectory normalized Levenshtein distance between the post-LLM-refinement operation sequence and the final human-annotated trajectory ($n{=}52$). Each operation is treated as an indivisible token. All 52 trajectories were modified by annotators; mean = 0.69, median = 0.77, with 19 (37\%) near-completely rewritten.
  }
  \label{fig:edit-distance}
\end{figure}



\section{Classical Model Details} \label{app:classical-solvers}

Alongside the LLM and SLM solvers, we evaluate four classical sequence-prediction baselines---\textsc{$n$-gram}, \textsc{online $n$-gram}, \textsc{lstm}, and \textsc{xgboost}.
All four share a common featurization pipeline and a common decoding loop; they only differ in how they map a window of past operations to the next predicted operation.
This section documents the shared infrastructure and the model-specific design choices.

\paragraph{Shared featurizer.}
Every operation in the trajectory is parsed once into a structured record with five primary fields: \textit{op-type} (one of $\sim 28$ symbolic operation types such as \op{INPUT}, \op{FONT\_BOLD}, \op{BORDER\_ALL}), the range geometry (\textit{start row}, \textit{start col}, \textit{height}, \textit{width}), and a \textit{value-type} bucket (e.g., \op{number}, \op{string}, \op{color}, \op{align\_h}; 18 buckets total).
For range-valued features we support two equivalent representations.
The \textbf{absolute} mode keeps the raw geometry $(\text{row}, \text{col}, h, w)$.
The \textbf{relative} mode replaces these with deltas with respect to the previous operation $(\Delta\text{row}, \Delta\text{col}, \Delta h, \Delta w)$ and additionally tags the move with one of 11 \textit{movement classes} (\op{same}, \op{right1}, \op{down\_n}, \op{diagonal}, \op{sheet\_change}, \ldots).
The featurizer is incremental: new operations are appended as they arrive, and a history rewind triggers a full re-featurization.
At prediction time, the same featurizer also reconstructs symbolic strings from predicted feature tuples, picking a sensible default value (e.g., \op{Calibri} for \op{FONT\_NAME}, \op{Thin, Continuous, \#000000} for borders) when only a value-type is predicted.


\paragraph{\textsc{Trained $n$-gram}.}
This baseline builds back-off frequency tables from the training trajectories.
The abstract token used as an $n$-gram key is the tuple $(\textit{op-type}, h, w, \textit{value-type})$ in absolute mode and $(\textit{op-type}, \textit{movement-class}, \textit{value-type})$ in relative mode---deliberately coarser than the full symbolic string, so that statistical regularities (``bold is usually followed by fill on the same range'') survive sparsity.
At inference time, we look up the longest matching context (up to order $n{=}5$ by default), back off to shorter contexts as needed, and pick the most frequent next token.
Range geometry is then predicted separately: in absolute mode by extrapolating an arithmetic progression over recent operations of the same type; in relative mode by sampling the most-common delta seen for that abstract pair during training.

\paragraph{\textsc{Online $n$-gram}.}
The online variant ignores the training corpus entirely and instead learns the statistics \emph{of the current trajectory} as it unfolds.
Its primary mechanism is a hash-indexed \textbf{suffix index}: after each new operation, every suffix of the history up to length $n_{\max}{=}5$ is hashed into a bucket, allowing $O(1)$ longest-match lookup.
At prediction time, the solver searches for the longest suffix of the current history that has occurred earlier; if a match of length $\geq 2$ is found, the operations that followed the previous occurrence are emitted as the predicted continuation.
This makes \textsc{online $n$-gram} especially strong at predicting locally repeating patterns (e.g., repeating a row of borders across a header band) without any external training data.
When no suffix match is available, the solver falls back to within-trajectory $n$-gram frequencies and arithmetic-progression range extrapolation, identical in spirit to the trained \textsc{$n$-gram}.

\paragraph{\textsc{LSTM/GRU}.}
The neural baseline is a small GRU (the \texttt{lstm} label is historical) trained on the same trajectories.
Each operation is tokenized into a 6-tuple: $(\textit{op-id}, r_1, r_2, r_3, r_4, \textit{value-type-id})$, where the four range fields are bucketed into 101 integer bins each---absolute row/col/height/width bins in absolute mode, or signed deltas clipped to $[-50, +50]$ in relative mode.
The model embeds each field separately, concatenates the embeddings, and feeds the resulting sequence through a 2-layer GRU (hidden size 128) with five output heads: one classifier over op-types, four classifiers over range bins, and one classifier over value-types.
Decoding uses a fixed 32-step context window and proceeds auto-regressively through the shared decoding loop; predicted bins are un-bucketed and combined with a default value from the value-type table to reconstruct a symbolic operation.

\paragraph{\textsc{XGBoost}.}
The tree-based baseline uses the same featurized window of the last $K{=}10$ operations and trains five gradient-boosted models in parallel: a multi-class classifier for the next op-type and four regressors (one per range coordinate, either absolute or relative).
Predicted ranges are clipped to valid ranges and discretized back to integer cell coordinates; the predicted op-type is mapped to a default value via a learned value-lookup table.
Like the GRU, the XGBoost solver plugs into the shared decoder and benefits from the same stopping rules.

\paragraph{Why these baselines.}
The four solvers cover a deliberate spectrum: \textsc{online $n$-gram} captures \emph{purely local} repetition with zero training; \textsc{$n$-gram} captures \emph{global frequency} of short patterns learned from the corpus; \textsc{xgboost} learns \emph{tabular feature interactions} over a fixed window; and \textsc{lstm} learns \emph{distributed sequence representations} that can in principle model longer-range structure.
Together they provide a low-cost, non-LLM reference point against which the language-model solvers in Section~\ref{sec:experiments} can be calibrated.

\section{Fine-tuning Setup}\label{app:slm-training}

We fine-tune two SmolLM2 variants (135M and 360M parameters) on synthetically generated
operation sequences using a causal language-modelling objective.
The training corpus is produced by the same symbolic + LLM pipeline used for the benchmark
(Section~\ref{sec:benchmark}), applied to a separate pool of 11{,}997 publicly available
workbooks disjoint from the evaluation set.
Each training example is a contiguous slice of a trajectory; the model sees up to 32 prior
operations as context and is trained to predict the next 16 operations.

\paragraph{Pre-processing.}
The text representation matches the evaluation solver: each operation is the pipe-delimited
symbolic form (Section~\ref{sec:setup}), sheet names are stripped (the active sheet is
provided implicitly), and long cell values are shortened to 32 characters with 2$\times$2
corner cells preserved for 2D arrays.
Training examples are sampled with a stride of 16 operations to balance coverage and
diversity, and we filter to the 99th percentile of sequence length to drop a small number
of outliers.

\paragraph{Prompt template.}
Each training example is wrapped in a minimal prompt to match the inference-time format:
\begin{quote}\itshape\small
Complete the sequence of actions to build the following spreadsheet by identifying and
extending key patterns.\\[2pt]
\{actions\}
\end{quote}
The model is trained to continue from the prompt by emitting the next operations, one per
line.

\paragraph{Training hyperparameters.}
Both variants use AdamW with a cosine schedule, weight decay $0.01$, gradient checkpointing,
and fp16 mixed precision.
We use a batch size of 4 with 8 gradient-accumulation steps (effective batch 32), a
learning rate of $8\!\times\!10^{-5}$ with 8\% warmup, and a maximum sequence length of
2048 tokens.
The 360M model is trained for 6 epochs and the 135M model for 5; checkpoints are saved
every 1{,}000 steps and the lowest-eval-loss checkpoint is used for evaluation.
A 5\% held-out split monitors evaluation loss, and a fixed set of precision-1 test cases
monitors generation accuracy throughout training.
We do \emph{not} use LoRA---all parameters are updated.
Training takes roughly two days for the 360M model on a single T4 GPU and about a day for
the 135M variant.

\section{Solver Prompts}\label{app:prompts}

All zero-shot LLM solvers share a single Jinja2-templated system prompt.
The two evaluation settings (single-action repredict vs.\ multi-action prediction) are
switched by the boolean \texttt{single}, set to \texttt{(num\_op\_to\_pred == 1)} at render
time; the rest of the template is identical.
Figure~\ref{fig:solver-prompt} shows the template in full, with the lengthy operation-syntax
reference abridged for space (it is reproduced verbatim in the released code).
The user prompt is small: it lists the active sheet name (when applicable) followed by the
recent operation history.

\begin{figure*}[t]
\centering
\begin{lstlisting}[style=pseudo,frame=single,basicstyle=\scriptsize\ttfamily,
                   numbers=none,showstringspaces=false,language={}]
{%- set single = (num_op_to_pred == 1) -%}
You are an autocomplete copilot for spreadsheet authors.
{% if single %}Given the user's recent editing history, predict the single most likely next
operation they will perform.{% else %}Continue the user's current workflow without
inventing new, unrelated workstreams.{% endif %}

Decision rubric:
1. Infer the immediate intent from the latest few steps (table build, formatting sweep,
   etc.).{% if not single %} Finish that chunk before starting anything new.{% endif %}
2. Keep ranges tight and contiguous unless the history explicitly shows a jump to a
   different area.
3. Only suggest {% if single %}an operation that clearly advances{% else %}operations that
   clearly advance{% endif %} the current goal. If you cannot justify
   {% if single %}it{% else %}an action{% endif %} with the visible history, omit it.
{%- if not single %}
4. Never repeat large historical blocks verbatim - mirror the pattern, not the entire plan.
{%- endif %}
{{ '4' if single else '5' }}. Stop as soon as the workflow looks complete or uncertain.

Output contract:
- {% if single %}Emit exactly one well-justified, high-confidence operation{% else %}Emit
  only well-justified, high-confidence steps{% endif %} in the format:
    OPERATION | RANGE | VALUE
- Ranges do not include sheet names (the active sheet is provided separately).
- Prefer formulas with relative references that match the user's existing convention.
{%- if num_op_to_pred %}
- Do not exceed {{ num_op_to_pred }} total operations. Returning fewer (including zero)
  is acceptable when intent is ambiguous.
{%- endif %}

Available operations (use EXACT names, no wildcards):
  Data:       INPUT, PASTE_FROM, AUTOFILL
  Formatting: NUMBER_FORMAT, FILL_COLOR
  Font:       FONT_BOLD, FONT_ITALIC, FONT_SIZE, FONT_COLOR, FONT_UNDERLINE, FONT_NAME
  Alignment:  ALIGN_HORIZONTAL, ALIGN_VERTICAL
  Borders:    BORDER_LEFT, BORDER_RIGHT, BORDER_TOP, BORDER_BOTTOM, BORDER_OUTSIDE,
              BORDER_ALL, BORDER_INSIDE_HORIZONTAL, BORDER_INSIDE_VERTICAL
  Other:      MERGE, UNMERGE, WRAP_TEXT, TEXT_ORIENTATION

<...value-format rules per operation type (INPUT, FILL_COLOR, FONT_*, ALIGN_*,
   NUMBER_FORMAT, BORDER_*, ...); see released code for full text...>

Rules for INPUT:
- Single cell:    INPUT | A1 | "Hello World"
- Multiple cells: INPUT | A1:C2 | [["a","b","c"],["d","e","f"]]
  The 2D array dimensions MUST exactly match the range (rows x cols).

Rules for PASTE_FROM:  PASTE_FROM | destination | source_range | mode
   Modes: all, values, formats, formulas    e.g.   PASTE_FROM | A10:C12 | A1:C3 | all

Rules for AUTOFILL:    AUTOFILL | destination_range | source_range
   The destination must fully contain the source and extend on exactly one axis.
   e.g.   AUTOFILL | A1:A10 | A1:A3

Clearing: use "clear" for INPUT / FILL_COLOR / BORDER_*. For format resets, set the Excel
default directly (FONT_BOLD -> false, NUMBER_FORMAT -> General).

Note: history may abbreviate large arrays with "..."; do NOT use this marker in
predictions - always emit the full, exact values.

Example syntax:
  INPUT       | A1      | "Hello World"
  INPUT       | A1:B2   | [["a","b"],["c","d"]]
  FILL_COLOR  | A1:B5   | #FF0000
  BORDER_ALL  | A1:C3   | Thin, Continuous, #000000
  FONT_BOLD   | A1      | true
  MERGE       | A1:C1   | true
  PASTE_FROM  | A10:C12 | A1:C3 | all
  AUTOFILL    | A1:A10  | A1:A3
\end{lstlisting}
\caption{Jinja2 system-prompt template used by all zero-shot LLM solvers. The
\texttt{single} flag (\texttt{true} iff \texttt{num\_op\_to\_pred == 1}) switches between
the single-action repredict and multi-action variants; the per-call cap line is included
only when \texttt{num\_op\_to\_pred} is non-null. The full value-format rules per operation
type are abridged here for space; the complete template is reproduced verbatim in the
released code.}
\label{fig:solver-prompt}
\end{figure*}

\section{Variance and Reproducibility}\label{app:variance}

To quantify run-to-run variance, we re-ran the default GPT-5 configuration four
additional times in each of the two evaluation settings, giving five runs per setting.
The solver is invoked at temperature 0, but small non-determinism remains in the chat API
backend; the prompt, hyperparameters, and the seed of the evaluation loop itself are held
fixed across reruns.
Table~\ref{tab:variance} reports per-run \textsc{uas} and \textsc{ar} together with their
mean and standard deviation.

Across both settings the spread is small: \textsc{uas} varies by under one point in the
single-action setting and by under two points in the multi-action setting (where the model
is free to emit longer, more variable continuations); acceptance rate is even tighter.
We therefore report a single representative run elsewhere in the paper.

\begin{table}[h]
  \caption{Five-run variance for the default GPT-5 configuration ($s{=}1$,
  $c{=}32$, \textsc{greedy}). ``Run 0'' is the run reported as the base in the main paper.}
  \label{tab:variance}
  \centering
  \small
  \begin{tabular}{llrr}
    \toprule
    \textsc{setting} & \textsc{run} & \textsc{uas} & \textsc{ar} \\
    \midrule
    \multirow{6}{*}{Multi-action}
      & Run 0   & 22.31 & 19.55 \\
      & Run 1  & 23.29 & 19.87 \\
      & Run 2  & 23.32 & 20.11 \\
      & Run 3  & 23.72 & 20.52 \\
      & Run 4  & 24.04 & 19.92 \\
      \cmidrule(lr){2-4}
      & Mean $\pm$ s.d.\ & 23.34 $\pm$ 0.65 & 19.99 $\pm$ 0.36 \\
    \midrule
    \multirow{6}{*}{Single-action repredict}
      & Run 0   & 27.35 & 31.73 \\
      & Run 1  & 27.42 & 31.43 \\
      & Run 2  & 26.49 & 31.15 \\
      & Run 3  & 26.59 & 31.01 \\
      & Run 4  & 27.11 & 31.40 \\
      \cmidrule(lr){2-4}
      & Mean $\pm$ s.d.\ & 26.99 $\pm$ 0.43 & 31.34 $\pm$ 0.28 \\
    \bottomrule
  \end{tabular}
\end{table}

\section{Some more insights}

\begin{figure}[t]
\begin{floatrow}
\ffigbox[.475\textwidth]{%
  \includegraphics[width=\linewidth]{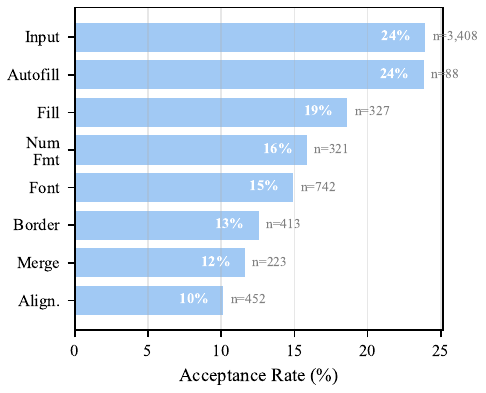}
}{%
  \caption{Acceptance rate by last context operation category (uses default settings). Predictions following Input operations are most likely to be accepted.}%
  \label{fig:last_op_acceptance}
}%
\ffigbox[.475\textwidth]{%
  \includegraphics[width=\linewidth]{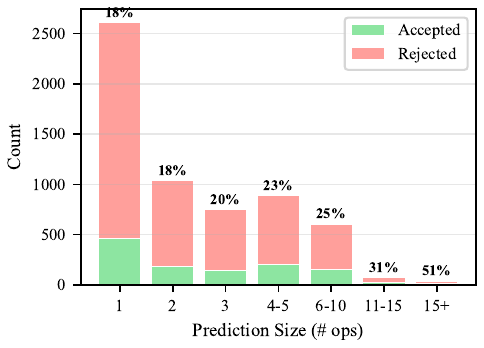}
}{%
  \caption{Distribution of predictions by number of operations (uses default settings). Bars show accepted (green) and rejected (red) prediction counts. Percentages indicate acceptance rate for each size bin.}%
  \label{fig:prediction-volume-size}
}%
\end{floatrow}
\end{figure}

\paragraph{Last operation matters.}
The last operation performed by the user before a prediction is triggered has a clear effect on acceptance in multi-action prediction setting (Figure~\ref{fig:last_op_acceptance}).
Predictions following \op{input} or \op{autofill} operations achieve the highest acceptance rates ($\sim$24\%), roughly twice that of predictions following \textsc{merge} or \textsc{align} ($\sim$10--12\%). However, \op{autofill} is predicted much less often.
This suggests the model predicts more reliably during data-entry phases with predictable sequential patterns, while operations that signal workflow transitions are harder to anticipate.

\begin{figure}
\begin{floatrow}
\ffigbox[.475\textwidth]{%
  \includegraphics[width=0.4\textwidth]{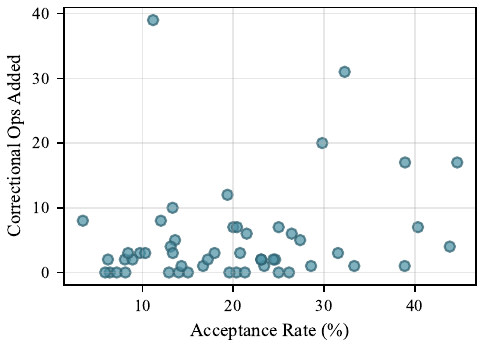}
}{%
  \caption{Relationship between acceptance rate and correctional operations per file (uses default settings). Each point represents one spreadsheet. Files with higher acceptance rates tend to have fewer correctional operations, suggesting that users who accept more predictions make fewer manual corrections.}\label{fig:correctional-vs-acceptance}
}%
\ffigbox[.475\textwidth]{%
  \includegraphics[width=0.4\textwidth]{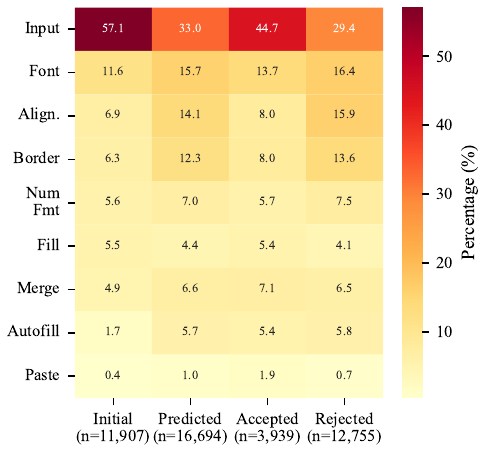}
}{%
  \caption{Operation category distribution across different sequence types (uses default settings). Columns show the percentage breakdown by category for initial (target) sequences, final user states, predicted operations, accepted predictions, and rejected predictions.}\label{fig:category-heatmap}
}%
\end{floatrow}
\end{figure}

\paragraph{Corrections are bounded and don't grow with acceptance.}
Figure~\ref{fig:correctional-vs-acceptance} plots correctional operations (user actions that undo false positives) against acceptance rate per file.
The vast majority of files require fewer than 10 corrections regardless of acceptance rate, with no monotonic relationship between the two: even files with $>$30\% acceptance rarely exceed a handful of corrections, while a few outliers across the acceptance spectrum incur 20+ corrections.
This suggests the cost of false positives is small in aggregate and largely orthogonal to how much the user accepts---an encouraging signal for predictive autocompletion in this domain.

\paragraph{Category distribution reveals prediction bias.}
Comparing the \emph{Initial} and \emph{Predicted} columns in Figure~\ref{fig:category-heatmap} exposes a systematic shift: the model under-predicts \op{input} (33.0\% predicted vs.\ 57.1\% in ground truth) and over-predicts formatting categories such as \op{align} (14.1\% vs.\ 6.9\%) and \op{border} (12.3\% vs.\ 6.3\%).
This bias correlates with rejection---\op{align} accounts for 15.9\% of rejected but only 8.0\% of accepted predictions, and \op{border} shows the same pattern (13.6\% vs.\ 8.0\%)---so the predictor's distribution amplifies its weaknesses.
Future work could explore category-aware prediction constraints or weighted training objectives to correct this imbalance.


\paragraph{Larger predictions are accepted more often---when they are made.}
Figure~\ref{fig:prediction-volume-size} shows that acceptance rate rises monotonically with prediction length: single-op predictions are accepted only 18\% of the time, while predictions of 11--15 ops reach 31\% and the rare 15+-op bin reaches 51\%.
The volume distribution is inversely shaped---short predictions dominate the count---so most rejections come from small predictions that fail to clear the utility threshold of the acceptance heuristic.
When the model commits to a longer prediction it is also more likely to have recognised a strong, repeating pattern (e.g., propagating a header style across a row), which is exactly the regime in which our heuristic accepts.
This indicates that prediction \emph{confidence} (implicit in the model's willingness to emit a longer sequence) is more predictive of acceptance than prediction length itself.



\section{Visualizing trajectories}
\begin{figure*}[t]
    \centering
    \includegraphics[width=\textwidth]{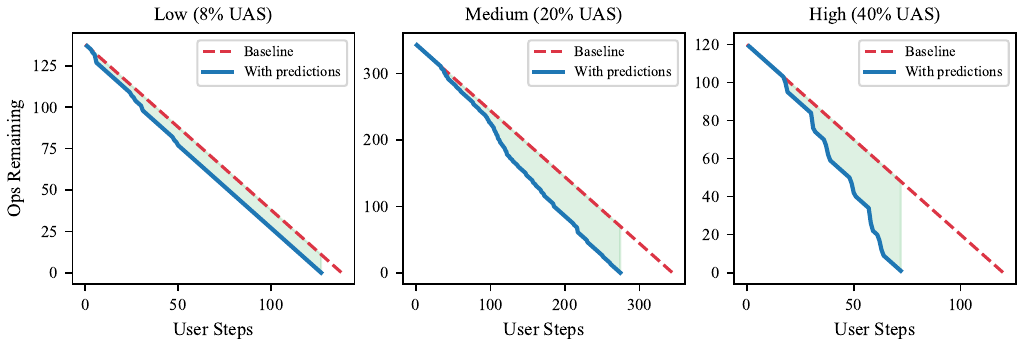}
    \caption{Progress curves for three representative files showing operations remaining versus user steps (uses default settings). The blue line shows actual progress with model assistance, while the red dashed line represents the manual baseline (no predictions). The shaded green area represents user effort saved. Percentages indicate total operations saved compared to manual completion.}
    \label{fig:progress-curves}
\end{figure*}

\begin{figure*}[t]
    \centering
    \includegraphics[width=\textwidth]{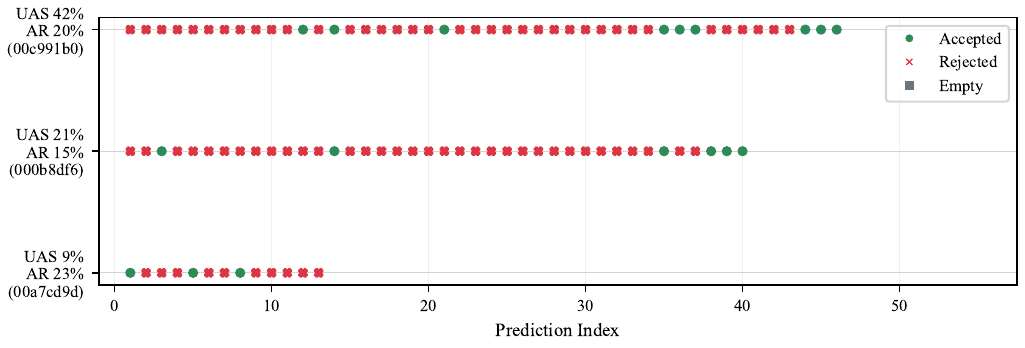}
    \caption{Prediction timelines for three files with varying acceptance rates (uses default settings). Each marker represents a prediction: accepted (circles), rejected (X), or empty (squares). Files with higher acceptance rates show denser clusters of accepted predictions.}
    \label{fig:prediction-timeline}
\end{figure*}

Figures~\ref{fig:progress-curves} and~\ref{fig:prediction-timeline} provide complementary views of system behavior. The progress curves show cumulative impact --- how accepted predictions (green markers) accelerate progress relative to the manual baseline --- while the prediction timeline reveals temporal patterns, showing when predictions succeed or fail throughout the trajectory.


\section{Extended example}

\begin{figure*}[tb]
\centering
\small

\definecolor{undocolor}{RGB}{180, 80, 80}      
\definecolor{synthcolor}{RGB}{60, 120, 160}    
\definecolor{fixcolor}{RGB}{120, 90, 150}      
\definecolor{tpcolor}{RGB}{80, 140, 80}        
\definecolor{notecolor}{RGB}{120, 120, 120}    
\definecolor{panelbg}{RGB}{248, 248, 252}      
\definecolor{triggercolor}{RGB}{100, 100, 180} 
\definecolor{acceptbg}{RGB}{230, 245, 230}     
\definecolor{rejectbg}{RGB}{255, 240, 240}     
\definecolor{sectionbg}{RGB}{240, 245, 250}    

\newcommand{\trigger}{{\color{triggercolor}\scriptsize$\blacktriangleright$~\textit{trigger}}}
\newcommand{\sectionhead}[1]{\cellcolor{sectionbg}\textit{#1}}

\begin{tabular}{@{}p{0.46\textwidth}@{\hspace{0.02\textwidth}}|@{\hspace{0.02\textwidth}}p{0.48\textwidth}@{}}

\begin{tabular}{@{}rl@{}}
\multicolumn{2}{@{}l}{\fcolorbox{triggercolor}{panelbg}{\textbf{~(a) Online Evaluation Trace~}} {\small (interval $s{=}3$)}}\\[0.6em]

\op{8.} & \op{BOLD A5:E5 True} \\
\op{9.} & \op{FILL A5:E5 Blue} \\
\op{10.} & \op{BORDER A5:E5 Thin} \hfill \trigger\\[0.3em]
\cline{1-2}\\[-0.7em]
\multicolumn{2}{@{}l}{%
  \hspace{0.1em}\colorbox{rejectbg}{\fbox{\parbox{0.40\textwidth}{\small%
  \textbf{Prediction} $\hat{A}_{10}$:\\[0.2em]
  \op{~~VAL A6 "Redesign", VAL B6 "Alice",}\\
  \op{~~FILL A6:B7 Green, VAL C6 "Dec 20",}\\
  \op{~~BORDER A6:E6 Thin}\\[0.4em]
  \colorbox{white}{\strut Prec=33\%, \textsc{uas}=0}~~$\Rightarrow$~~\textcolor{red!70!black}{\textbf{Reject}}
  }}}%
}\\[0.6em]
\multicolumn{2}{@{}l}{{\color{notecolor}\small\itshape~~User rejects and continues with ground truth action}}\\[0.6em]

\op{11.} & \op{VAL A6 "Redesign"} \\
\op{12.} & \op{VAL B6 "Alice"} \\
\op{13.} & \op{FILL A6:D6 Green} \hfill \trigger\\[0.3em]
\cline{1-2}\\[-0.7em]
\multicolumn{2}{@{}l}{%
  \hspace{0.1em}\colorbox{acceptbg}{\fbox{\parbox{0.40\textwidth}{\small%
  \textbf{Prediction} $\hat{A}_{13}$:\\[0.2em]
  \op{~~VAL C6 "Dec 15", VAL D6 "Complete"}\\[0.4em]
  \colorbox{white}{\strut Prec=100\%, \textsc{uas}=+2}~~$\Rightarrow$~~\textcolor{green!50!black}{\textbf{Accept}}
  }}}%
}\\[0.6em]
\multicolumn{2}{@{}l}{{\color{notecolor}\small\itshape~~$\hookrightarrow$ User accepts; Effects in-place, Ground truth adapted}}\\[0.6em]

\op{14.} & \op{VAL A7 "Mobile App"} \hfill {\small\color{notecolor}(continues...)}\\
\end{tabular}

&

\begin{tabular}{@{}lll@{}}
\multicolumn{3}{@{}l}{\fcolorbox{triggercolor}{panelbg}{\textbf{~(b) Prediction Analysis~}} {\small (detail of rejected $\hat{A}_{10}$)}}\\[0.6em]

\multicolumn{3}{@{}l}{\colorbox{sectionbg}{\strut\textit{~Future ground truth $F_{>10}$ (remaining ops):~}}} \\[0.3em]
\op{F1.} & \op{VAL A6 "Redesign"} & \\
\op{F2.} & \op{VAL B6 "Alice"} & \\
\op{F3.} & \op{FILL A6:D6 Green} & \\
\op{F4.} & \op{VAL C6 "Dec 15"} & \\
\op{F5.} & \op{VAL D6 "Complete"} & \\[0.5em]

\multicolumn{3}{@{}l}{\colorbox{sectionbg}{\strut\textit{~Prediction $\hat{A}_{10}$ with classification:~}}} \\[0.3em]
\op{P1.} & \op{VAL A6 "Redesign"} & \colorbox{tpcolor!15}{{\color{tpcolor}\checkmark~TP (1)}} \\
\op{P2.} & \op{VAL B6 "Alice"} & \colorbox{tpcolor!15}{{\color{tpcolor}\checkmark~TP (1)}} \\
\op{P3.} & \op{FILL A6:B7 Green} & \colorbox{synthcolor!15}{{\color{synthcolor}$\diamond$~partial (2 TP, 2 FP)}} \\
\op{P4.} & \op{VAL C6 "Dec 20"} & \colorbox{fixcolor!15}{{\color{fixcolor}$\neq$~MM (1)}} \\
\op{P5.} & \op{BORDER A6:E6 Thin} & \colorbox{undocolor!15}{{\color{undocolor}$\times$~FP (5)}} \\[0.4em]

\multicolumn{3}{@{}l}{%
\small
\fcolorbox{gray!30}{white}{%
\textcolor{tpcolor}{\textbf{TP}}=4~~%
\textcolor{undocolor}{\textbf{FP}}=7~~%
\textcolor{fixcolor}{\textbf{MM}}=1%
\hspace{1em}$\Rightarrow$~~\textbf{Prec}=33\%~~\textbf{\textsc{uas}}=0%
}}\\[0.7em]

\multicolumn{3}{@{}l}{\colorbox{sectionbg}{\strut\textit{~Adapted $F'_{>10}$ (if $\hat{A}_{10}$ accepted):~}}} \\[0.3em]
{\color{undocolor}\op{1'.}} & {\color{undocolor}\op{CLEAR BORDER A6:E6}} & {\color{notecolor}\small\textit{$\leftarrow$ undo P5 (FP)}} \\
{\color{undocolor}\op{2'.}} & {\color{undocolor}\op{CLEAR FILL A7:B7}} & {\color{notecolor}\small\textit{$\leftarrow$ undo P3 partial (FP)}} \\
{\color{synthcolor}\op{3'.}} & {\color{synthcolor}\op{FILL C6:D6 Green}} & {\color{notecolor}\small\textit{$\leftarrow$ synth from F3 (FN)}} \\
{\color{fixcolor}\op{4'.}} & {\color{fixcolor}\op{VAL C6 "Dec 15"}} & {\color{notecolor}\small\textit{$\leftarrow$ fix P4 (MM)}} \\
{\color{synthcolor}\op{5'.}} & {\color{synthcolor}\op{VAL D6 "Complete"}} & {\color{notecolor}\small\textit{$\leftarrow$ synth F5 (FN)}} \\[0.1em]
\multicolumn{3}{@{}r@{}}{{\color{notecolor}\small\textit{F1--F2 covered by P1--P2 (TP)}}} \\[0.4em]
\multicolumn{3}{@{}l}{\fcolorbox{gray!40}{rejectbg}{\small\textbf{Net:} 5 $\to$ 5 ops (no benefit $\Rightarrow$ rejected)}}\\
\end{tabular}

\end{tabular}

\vspace{0.8em}

\fcolorbox{triggercolor}{panelbg}{\textbf{~(c) Visual Comparison~}} {\small (spreadsheet states at $t{=}10$)}\\[0.6em]
\begin{tabular}{@{}c@{\hspace{1.5em}}c@{\hspace{1.5em}}c@{}}
\fbox{\includegraphics[width=0.30\textwidth]{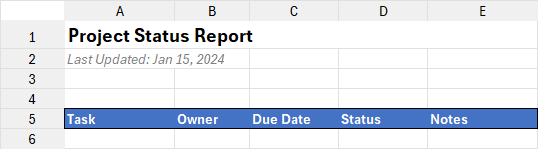}} &
\fbox{\includegraphics[width=0.30\textwidth]{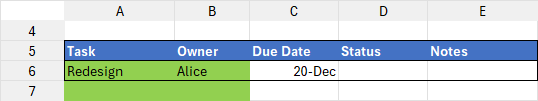}} &
\fbox{\includegraphics[width=0.30\textwidth]{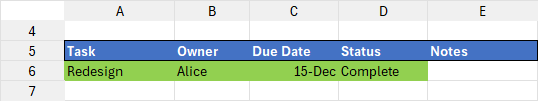}} \\[0.4em]
{\small\textit{Before $\hat{A}_{10}$ (state at $t{=}10$)}} &
{\small\color{undocolor}\textit{After $\hat{A}_{10}$ (with errors)}} &
{\small\color{tpcolor}\textit{Ground Truth (correct)}}
\end{tabular}

\caption{Online evaluation with detailed prediction analysis. (a)~Evaluation trace showing two trigger points: at $t{=}10$ the prediction $\hat{A}_{10}$ is rejected (low precision), at $t{=}13$ prediction $\hat{A}_{13}$ is accepted (perfect precision, saves 2 ops). (b)~Detailed breakdown of the rejected prediction $\hat{A}_{10}$: each (cell, property) classified as TP, FP, FN, or MM. If accepted, ground truth adapts: {\color{undocolor}inverse ops} undo FPs; {\color{synthcolor}synthesized ops} fill FNs; {\color{fixcolor}corrections} fix MMs. (c)~Visual comparison of spreadsheet states.}
\label{fig:eval_metrics}
\end{figure*}



\end{document}